\documentclass{cimento}

\usepackage{graphicx}

\title{An Overview of the {\sl SPACE} Mission Proposal}
\author{M.~Robberto\from{ins:x1}, A.~Cimatti\from{ins:x2} and the {\sl SPACE} science team\thanks{The complete list of the {\sl SPACE}
science team members is available at www.spacesat.info}}
\instlist{\inst{ins:x1} Space Telescope Science Institute, Baltimore, MD, 21218, USA (robberto@stsci.edu)
  \inst{ins:x2} Dipartimento di Astronomia, Universit\'a di Bologna, Italy (a.cimatti@unibo.it)}

\begin{document}

\maketitle

\begin{abstract}
{\sl SPACE} ({\sl SP}ectroscopic {\sl A}ll-sky {\sl C}osmic {\sl E}xplorer) is a class-M mission proposed to ESA for the Cosmic Vision 2015-2025 call and recently promoted to the next assessment study phase. {\sl SPACE} will produce the first all-sky spectroscopic survey of the Universe, taking spectra of more than 500 million galaxies over a wide range of redshifts. {\sl SPACE} will operate in slit mode (MEMS) at $R\sim400$ between $0.8$ and $1.8\mu$m down to $AB\sim 23$, providing redshifts to an accuracy of $\Delta z \sim 0.001$, regardless on the presence of bright emission lines, together with the most relevant physical and evolutionary properties. The catalog of spectroscopic redshift will allow to place the ultimate constraints on the Baryon Acoustic Oscillations and the nature of Dark Energy. By obtaining the first 3-D all-sky map of the Universe at $z\sim 2$ and beyond, {\sl SPACE} will trace the growth rate of cosmic structures, the large scale structure of luminous baryons and its cosmic evolution. Besides the all-sky survey, {\sl SPACE} will carry out a deep extragalactic survey over an area of $\simeq 10$ sq. deg., enabling a most powerful supernova search program, and a galactic plane survey in integral field mode. Approximately 30\% of the time will be open.  
Due to its  versatility, {\sl SPACE} is a ``self-sufficient'' observatory which can attack and solve the most compelling questions on the nature of the Dark Energy without complementary data from the Earth or space. Its unique wide-field capabilities in the near-IR make {\sl SPACE} the ideal complement to JWST, ALMA, and the future 25-50 m telescopes.
\end{abstract}

\section{Introduction}
All-sky surveys represent the workhorses of astronomy. Following the continuous advances in detector technology, astronomers in the last decades have imaged the sky with increasing resolution and sensitivity over several decades of the electromagnetic spectrum. However, a deep {\sl spectroscopic} survey of the Universe has never been done. To collect hundreds
of millions of spectra of faint galaxies is a daunting task, posing extreme requirements on the multiplexing capability of the instrumentation and on the
integration time with the largest telescopes. From the Earth surface the situation appears especially hopeless in the near-IR, the most important wavelength range for cosmological studies, due to the 
atmospheric background and opacity gaps which limit the sensitivity and prevent
continuous observations of the main spectral features at various redshifts. The common
wisdom is that these observations are presently impossibly expensive\cite{REF:Peacock} and therefore 
near-IR multiband photometry is the only way of obtaining photometric redshifts for large samples of galaxies. 
``Photo-z's'' still require formidable investments of telescope time for calibration, remain 
vulnerable to errors and have limited accuracy.

Yet the scientific return of an all-sky survey in the near-IR performed in
full spectroscopic mode would be enormous. A catalog of spectral redshifts 
for hundreds of million of galaxies would allow to build the 3-d atlas of the universe. 
The distribution of galaxies would allow measuring the Baryonic Acoustic 
Oscillations (BAOs) caused by sound waves propagating
through the primordial plasma in the early Universe and imprinted in the matter distribution. 
Detecting and measuring the BAOs at different redshifts one can
jointly constrain the geometry of the Universe and its content as a function of
redshift\cite{REF:Angulo}. The same dataset would simultaneously allow an accurate assessment of the 
growth of structures over the last 10 billion years, providing a  
method to discriminate between theories of dark energy and theories of modified gravity 
proposed to explain the acceleration of the universe \cite{REF:Guzzo, REF:Wang}. Besides redshifts, having the spectra
of millions of galaxies at redshift larger than $z\sim 1$ would enable unprecedented
investigations into the formation, evolution, and interaction of galaxies in the history of Universe,
providing a long lasting legacy to be data mined for the years to come. 

We have envisioned a space mission that can accomplish these goals and 
proposed it to the European Space Agency in response to the 
call for the First Planning Cycle of the Cosmic Vision 2015-2025 program. It is
called {\sl SPACE} (SPectroscopic All-sky
Cosmic Explorer) and is intended as a class-M mission led by the European Space Agency
in collaboration with the space agencies of the ESA member states and NASA. 

\section{The {\sl SPACE} mission}

{\sl SPACE} will produce the three-dimensional evolutionary map of the Universe over the past 10 billion years  by taking spectra of more than $0.5\times10^9$ galaxies over the $3\pi$~sr of sky unobscured by the Galaxy. {\sl SPACE} will operate in the near-IR (0.8-1.8$\mu$m) with resolving power $R=\lambda/\Delta\lambda\simeq 400$ and exploit MEMS devices to perform slit-spectroscopy. We have envisioned a core science program composed by:
\begin{enumerate}
\item
The {\sl SPACE} {\bf All-Sky Survey}, which will reach $AB\simeq 23$ with $SNR=5$ per resolution element, 
targeting approximately 6,000 galaxies simultaneously across a 0.4 square degree field of view. Since there is no input catalogue of sources at these faint magnitudes, at each pointing {\sl SPACE}
will preliminary perform H-band imaging down to $AB\simeq23$ for source selection, producing as a byproduct the deepest all-sky imaging survey ever in the near-IR. For each field {\sl SPACE} will automatically extract an unbiased, stellar mass selected sample of galaxies selecting approximately every third galaxy. Switching into a multi-slit spectroscopic mode, {\sl SPACE} will take near-IR spectra at magnitudes unreachable from the ground, fully exploiting the low celestial background at L2. Spectra will allow to  precisely locate ($\Delta z~\sim0.001$) each galaxy in space. In approximately 2.5 years {\sl SPACE} will cover $\sim 70\%$ of the sky detecting with unprecedented accuracy BAO patterns in the Universe between 5 to 10 billion years ago. By dividing the the sample of galaxy redshifts in slices of width $\Delta z\sim 0.5$
and measuring the BAO signature in each slice, {\sl SPACE} will achieve a 0.5\% accuracy in the BAO scale measurement from these redshift slices, far superior to that expected from any other planned survey, as illustrated by Figure 1. 
The all-sky survey will also allow measuring the growth of structures\cite{REF:Guzzo, REF:Wang}, and enable a  large variety of studies of objects selected regardless of their spectral properties (e.g. presence of bright emission lines). 
\item
A deep spectroscopic survey of a smaller 10 deg$^2$ area, targeting $\simeq2$ million galaxies over to AB=26 and $2<z<10$. The {\sl SPACE} {\bf Deep Survey} will discover an enormous number of primordial galaxies at very high redshifts, assembling samples from as far back as 500 million years after the Big Bang. It will reveal a large number of Ly-$\alpha$ emission galaxies, star forming galaxies, rare examples of passively evolving galaxies at the earliest epochs, and the earliest QSOs. It will address the formation and early evolution of galaxies from the epoch of "first light."  Timing the Deep Survey observations with periodic intervals, {\sl SPACE} could discover $\sim 2300$ Type Ia Supernovae taking their spectra with an efficiency an order of magnitude higher than SNAP, thanks to its large field of view.  
\item
Using MEMS devices {\sl SPACE} can operate in integral field mode (Hadamard transforms). We therefore envision a third core program aimed to perform an integral field {\bf Galaxy Survey} of a 1 degree strip centered around the Galactic plane at $AB\simeq20$, similar to 
the GLIMPSE survey of the Galaxy done with Spitzer at longer wavelengths, but deeper and in spectroscopy. 
\end{enumerate}
Besides these core science programs, approximately 1/3 of the 5 years lifetime of {\sl SPACE} will be
made available for GO programs. 

\section{{\sl SPACE} configuration}
The design of {\sl SPACE} exploits payload and spacecraft components and concepts that are largely heritage from previous missions (Hershel, Planck, NIRSPEC/JWST) with the exception of the MEMS devices, which in our case are micromirror devices instead of microshutters as in NIRSPEC. Tables
\ref{tab:SPACE Mission Summary} summarize the main technical facts of {\sl SPACE}.

\begin{table}
  \caption{{\sl SPACE} MISSION SUMMARY}
  \label{tab:SPACE Mission Summary}
  \begin{narrowtabular}{1cm}{ll}
    \hline
Telescope diameter	& 1.5m\\
Wavelength range	&0.8-1.8 $\mu$m ($0.6-1.8\mu$m possible)\\
Overall mass	&$\simeq1500$~kg\\
Orbit/Launcher	&L2/Soyuz\\
Launch date	&Mid 2017\\
Mission Duration&5 years\\
Total field of view	&$51'\times27'$ (0.4 sq. degree) \\
Total nr. of apertures	&8.8 million  \\
Aperture field of view	&$0.75''\times0.75''$ \\
Dispersing element	&Prism $R\sim400$\\
Detector technology	&HgCdTe 0.4-1.8µm, 2k$\times$2k \\
Nr. of detectors	&16 (4 mosaics of $2\times2$ chips) \\
QE 			&$\ge 75\%$ average  \\
Readout noise		&5e$^-$/multiple read \\
Observing modes		&Wide field imaging, slit, slitless \\
			&and integral field spectroscopy (Hadamard)  \\
  \end{narrowtabular}
\end{table}

\section{Conclusions}
{\sl SPACE} will have a wide impact on astronomy. It will make an enormous contribution by providing the highest precision measurement of BAOs, the definitive measurement of the power spectrum of density fluctuations and its turnover, the tightest constraints on the nature of the dark matter and gravity by measuring the evolution of the cosmic expansion rate and the growth rate of cosmic large-scale structure, the characterization of the large scale distribution of galaxies, the study of the Galaxy and stars, and virtually impact every field of astronomy. Its wide-field, high-sensitivity infrared spectroscopic capability will remain unique for the foreseeable future. The datasets from the {\sl SPACE} core and GO programs will represent a long lasting legacy that will be data mined for many years to come.


\begin{figure}
\begin{minipage}[b]{0.5\linewidth} 
\centering
\includegraphics[width=6cm]{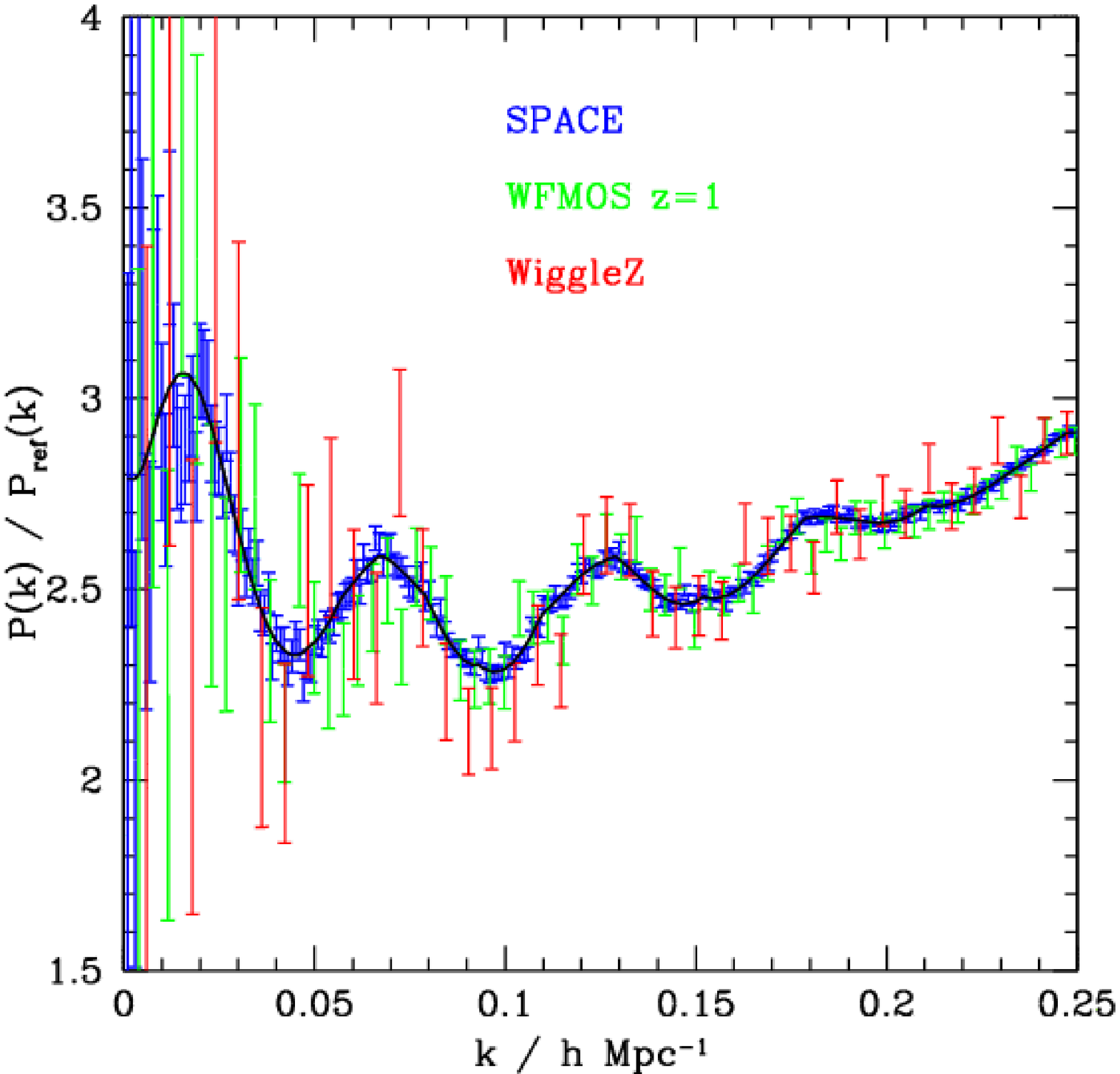}
\end{minipage}
\hspace{0.5cm} 
\begin{minipage}[b]{0.5\linewidth}
\centering
\includegraphics[width=6cm]{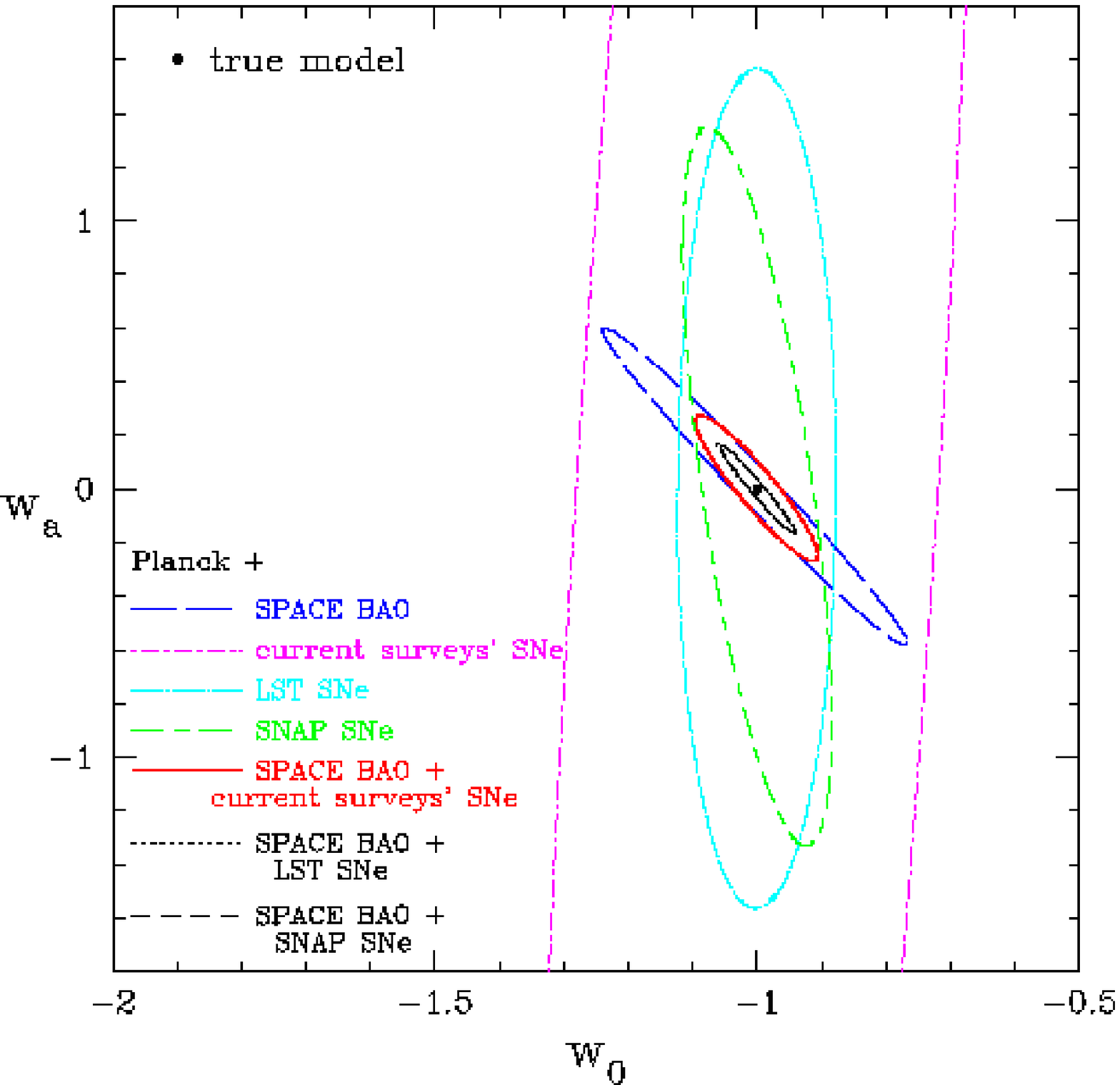}
\end{minipage}
\caption{{\sl Left}: the appearance of BAO in selected ongoing and future surveys. The measured power spectrum has been divided by a featureless reference spectrum to show the BAO more clearly. WiggleZ is an on-going survey of emission line selected galaxies at $z\sim0.7$. WFMOS is a spectroscopic, ground-based survey proposed for Subaru. The blue points show the measurements expected from the full survey volume of SPACE. The black solid line represents the theoretical model for the BAO. The high frequency sampling and small errors of P(k) from SPACE mean that we will achieve the definitive measurement of the BAO. The statistical power of the SPACE BAO measurement is around an order of magnitude better than that expected from WFMOS. {\sl Right}: The joint 68\% confidence interval constraints on the dark energy equation of state parameter (w$_0$) and its evolution with redshift (w$_a$). Several datasets can be combined to improve the constraints. SPACE will dramatically improve our knowledge of w$_0$ and w$_a$.}
\end{figure}



\begin{thebibliography}{0}
\bibitem{REF:Peacock} \BY{Peacock~J. et al.} 
  Report by the ESA-ESO Working Group on Fundamental Cosmology, astro-ph/0610906 (2006)
\bibitem{REF:Angulo} \BY{Angulo~R. et al.} 
Detectability of baryonic acoustic oscillations in future galaxy surveys, astro-ph/0702543 (2007)
\bibitem{REF:Guzzo} \BY{Guzzo~L. et al.} 
  Galaxy streaming motions as a probe of the origin of cosmic acceleration, Nature, submitted (2007)
\bibitem{REF:Wang} \BY{Wang~Y.} 
  Differentiating dark energy and modified gravity with galaxy redshift surveys, arXiv:0710.3885 [astro-ph] (2007)
\end{thebibliography}
\end{document}